\documentclass[twocolumn,superscriptaddress,aps,prb,floatfix]{revtex4-1}

\usepackage{graphicx}% Include figure files
\usepackage{dcolumn}% Align table columns on decimal point
\usepackage{bm}% bold math
\usepackage{color}
\usepackage[caption=false]{subfig} 

\usepackage{listings}
\usepackage{cancel}
\usepackage{ulem}

\definecolor{dkgreen}{rgb}{0,0.6,0}
\definecolor{gray}{rgb}{0.5,0.5,0.5}
\definecolor{mauve}{rgb}{0.58,0,0.82}

\usepackage{amsthm}
\usepackage{amsmath}
\usepackage{amssymb}

\newcommand{\Htc}{H_{\mathrm{TC}}}
\newcommand{\Hising}{H_{\mathrm{Ising}}}
\newcommand{\figref}[1]{Fig. \ref{#1}}

\newcommand{\te}{\tau_{\rm{esc}}}
\newcommand{\Tau}{\mathrm{T}}

\renewcommand*{\emph}[1]{{\it {#1}}}

\begin{document}

%\allowdisplaybreaks

\title{Stable quantum memories with limited measurement}
%\title{Finite temperature thresholds for stabilizer codes with few measurements}

\author{C. Daniel Freeman}
\email{daniel.freeman@berkeley.edu}
\affiliation{Berkeley Quantum Information \& Computation Center, University of California, Berkeley, CA 94720, USA}
\affiliation{Department of Physics, University of California, Berkeley, CA 94720, USA}

\author{Mohan Sarovar}
\email{mnsarov@sandia.gov}
\affiliation{Extreme-scale Data Science and Analytics, Sandia National Laboratories, Livermore, CA 94550, USA}

\author{C. M. Herdman}
\affiliation{Institute for Quantum Computing, University of Waterloo, Waterloo, ON N2L 3G1, Canada}
\affiliation{Department of Chemistry, University of Waterloo, Waterloo, ON N2L 3G1, Canada}
\affiliation{Department of Physics \& Astronomy, University of Waterloo, Waterloo, ON N2L 3G1, Canada}

\author{K. B. Whaley}
\affiliation{Berkeley Quantum Information \& Computation Center, University of California, Berkeley, CA 94720, USA}
\affiliation{Department of Chemistry, University of California, Berkeley, CA 94720, USA}

\date{\today}

\begin{abstract}
 We demonstrate the existence of a finite temperature threshold for a 1D stabilizer code under an error correcting protocol that requires only a fraction of the syndrome measurements.  Below the threshold temperature, encoded states have exponentially long lifetimes, as demonstrated by numerical and analytical arguments.  We sketch how this algorithm generalizes to higher dimensional stabilizer codes with string-like excitations, like the toric code. 
\end{abstract}

\maketitle

%%%%%%%%%%%%%%%%%%%%%%
%%%%%%%%%%%%%%%%%%%%%%
\section{Introduction}
\label{sec:Intro}
%%%%%%%%%%%%%%%%%%%%%%
 Quantum memories are an essential component for many quantum technologies, including quantum computing and quantum repeaters. In analogy to modern classical memories, one ideally wants a stable quantum memory that requires little or no active intervention and error correction. Unfortunately, no physical system that passively preserves quantum information indefinitely at finite temperatures and in an experimentally accessible number of dimensions is known \cite{Terhal2015ks}. Instead, the operation of all known practical quantum memories require a combination of passive elements (i.e. dissipative cooling) and active measurement and correction cycles to keep quantum information protected.  In this work, we study the degree to which the amount of active measurement and correction can be reduced while maintaining quantum memory stability (our notion of stability, to be quantified later, corresponds to exponentially long lifetime for encoded states below some finite threshold temperature). We develop a new decoding and correction protocol that enables one to trim the number of measurements to a fraction of the complete set of measurements normally considered, and still maintain quantum memory stability. 
 
 We restrict our attention to quantum memories defined through stabilizer codes. For near term architectures, stabilizer codes\cite{Gottesman98} have emerged as the leading candidate for encoding quantum information and subsequent active error correction in quantum hardware, with small scale architectures actively being developed and deployed\cite{Knill2001,Takita2017,Chiaverini2004,Corcoles2015}.  A tremendous amount of effort has gone into developing novel decoding and correction schemes for stabilizer codes, particularly the toric code.  Different schemes often emphasize different decoding features, like efficient decoding\cite{Kovalev2013,Wang2009,Duclos-Cianci2010}, locality\cite{Breuckmann2016,Torlai2016,Harrington2004,Herold2014}, robustness to particular sorts of noise\cite{Novais2013,Bombin2012,Wootton2015}, or use of dissipation\cite{Freeman2017,Hutter2012a,Hutter2014,Fujii2014,Pedrocchi2013,Pedrocchi2011,Chesi2010a,Chesi2010b,Hamma2009,Herdman2010, Young2012,Weimer2010, Dengis2014, Kapit2015, Bardyn2015}.
 
 In previous work \cite{Freeman2014}, we analyzed the finite temperature dynamics of the toric code, verifying the well-known no-go theorems for the upper bound to the lifetime of the toric code at finite temperature\cite{Bravyi2009, Alicki2009a, Chesi2010b, Yoshida2011b, Hastings2011a, Landon-Cardinal2013, Temme2014, temme2015fast}.  Using this analysis, we were able to construct a measurement-free protocol for protecting the encoded qubits of the toric code \cite{Freeman2017}, but these protocols again were limited by the no-go theorems, and only provided a multiplicative constant increase to the lifetime.  
 
 Building off this previous work, here we examine the extent to which a limited amount of measurement can increase the lifetime of stabilizer codes with string-like excitations.  In sum, we demonstrate an algorithm that, for any constant density of measurements for a stabilizer code with stringlike excitations undergoing dissipation at a fixed temperature, exhibits a threshold temperature, below which exponentially long lifetimes can be achieved in the encoded space.  The threshold temperature scales with the amount of measurement used---fewer measurements result in a smaller threshold temperature, whereas more complete measurement raises the threshold temperature.  This tradeoff is commensurate with and complements what is known about decoding the stabilizer codes in the presence of \emph{noisy}, but complete measurements\cite{Nickerson2016}.
 
 The remainder of the manuscript is structured as follows:  Section \ref{sec:stabfintemp} briefly reviews the theoretical tools used for performing simulation of stabilizer codes at finite temperature.  The content of this section is also expanded upon in refs \cite{Freeman2014,Freeman2017}.  Section \ref{sec:ecc_alg} includes the full description of our limited measurement algorithm, including a discussion of the expected low temperature error processes that cause the algorithm to fail, and a heuristic justification for the expectation of a threshold temperature below which a stable quantum memory is feasible.  Section \ref{sec:experiments} details our numerical investigations of our algorithm for the 1D Ising model.  Finally, Sec. \ref{sec:tc_alg} sketches how this algorithm could be generalized to higher dimensions, and Sec. \ref{sec:discussion} provides some concluding analysis and discussion.

 %%%%%%%%%%%%%%%%%%%%%%
\section{Stabilizer codes at Finite Temperature}
\label{sec:stabfintemp}

\subsection{Definitions}
\label{sec:Defs}
%%%%%%%%%%%%%%%%%%%%%%
In this section, we briefly review the theory of the 1D Ising model, as well as the Markovian open quantum systems formalism for evaluating its finite temperature dynamics.  The Hamiltonian for the 1D Ising model is

\begin{align}
\Hising = -\Delta \Sigma_i \sigma^i_z \sigma^{i+1}_z 
\label{eq:isingham}
\end{align}

\noindent where, for the remainder of the manuscript, unless explicitly stated otherwise, we assume $\Delta=1$.  This is exactly the Hamiltonian version of the repetition stabilizer \emph{code}\cite{Freeman2017}.  Note that the terms $\sigma^i_z \sigma^{i+1}_z$ correspond exactly to the parity check stabilizer operators of the repetition code (see Table \ref{tab:code_vs_hamiltonian}).

\begin{table}
\begin{center}
  \begin{tabular}{ l | r }
    \hline
    Repetition Stabilizer Code & 1d Ising Stabilizer Hamiltonian \\ \hline
    Encoded States & Ground states \\ \hline
    Bit flip errors & Excited states \\ \hline
    Decoding and error correction & Identical or by cooling \\ \hline
    \hline
  \end{tabular}
\end{center}
\caption{A short summary of the similarities and differences between the Ising model considered as a code (left panel) versus as a hamiltonian (right panel).}
\label{tab:code_vs_hamiltonian}
\end{table}

In the parlance of the 1D Ising model, bit flip errors are often also classified via the dual variables called \emph{domain walls} or defects.  Defects are simply locations on the 1D Ising chain where a stabilizer operator yields a measurement of $-1$---i.e., locations where neighboring spins point in different directions.  With periodic boundary, the number of these locations is always even, and a single bit flip event either creates a pair of such defects, deletes a pair of defects, or causes a defect to translate by one unit.  

As long as less than half the system has had errors, a majority rule decoder that has access to measurements of the full set of stabilizers ${\sigma^i_z \sigma^{i+1}_z}$ will reliably be able to correctly identify and remove errors.  When errors are completely independent (i.e., at very high temperature), we can define random variables $x_i=1$ when an error occurs on site $i$, and $0$ otherwise.  If these errors occur with probability $p$ on each spin, independently at random every error detection cycle, then Chernoff's bound gives an upper bound to the probability of an error in the encoded space, $P(\Sigma_i x_i \geq L/2) \leq \rm{exp}[-Lp\frac{\delta^2}{2+\delta}]$ for $\delta=1/2p-1$.  Thus, for complete measurement, errors in the encoded subspace are exponentially suppressed in system size, so long as the error rate is sufficiently small.

For much of the remainder of the manuscript, we consider how the decoding scheme changes when one does \emph{not} have access to the full set of stabilizer measurements.

Following Ref. \cite{Freeman2017}, we consider a simple local Ohmic, Markovian bath to model finite temperature effects.  This is modeled by the following master equation in Lindblad form:

\begin{align}
\dot{\rho }=\sum_{i}{2c_{i}\rho c^{\dagger }_i}-c^{\dagger }_{i}c_{i}\rho -\rho c^{\dagger }_{i}c_{i}, \label{eq:Lindblad}
\end{align}

\noindent Here $\rho$ is the density matrix, with Lindblad operators $c_{i}$ chosen to take the form:

\begin{equation}
\left \{ c_i (\Delta) \right \} = \left \{ \sqrt{\gamma(0)} T_{i}, \sqrt{\gamma(\Delta)} D^\dagger_{i}, \sqrt{\gamma(-\Delta)} D_{i} \right\} 
\end{equation}
where $T_{b}$ translates a defect by one unit, $D^\dagger_{b}$ creates a pair of defects, $D_{b}$ dissipates a pair of defects, and $\gamma(\cdot)$ is a rate function dependent on the details of the bath.  This bath is chosen to model the dynamics of local, single bit-flip errors.  In the Pauli basis, these operators take the following form.

\begin{align}
D^\dagger_{i}= &\frac{1}{4} \left(I \sigma_x I \right)\left(1+I \sigma_z \sigma_z\right)\left(1+ \sigma_z \sigma_z I\right) \notag \\
D_{i}= &\frac{1}{4} \left(I \sigma_x I \right)\left(1-I \sigma_z \sigma_z\right)\left(1-\sigma_z \sigma_z I\right) \notag \\
T_{i}= &\frac{1}{4} \left(I \sigma_x I \right)\left(1-I \sigma_z \sigma_z\right)\left(1+ \sigma_z \sigma_z I\right), \label{eq:LindbladDef}
\end{align}

\noindent By convention, we define $i$ to index the first qubit in these operators.

Finally, the remaining details of the bath are specified by the spectral density, which determines the rates with which the different Lindblad operators act:

\begin{align}
\gamma \left( \omega \right)=\xi \left  \vert \frac{\omega^n }{1-e^{-\beta \omega }} \right \vert \label{eq:gammadef}
\end{align}

\noindent where $n=1$ corresponds to an Ohmic spectral density, which is the choice we make for the remainder of the manuscript.  With this choice, in the absence of any error correcting protocol, it can be shown that the 1D Ising model has a system size independent thermal logical error rate given by\cite{Freeman2017}

\begin{align}
\Gamma_0 = \frac{\gamma(0)}{1+e^{1/T}} 
\end{align}

\noindent We define the bare lifetime of qubits evolving under the 1D Ising model Hamiltonian in contat with an Ohmic thermal bath to be $\Gamma_0^{-1}$.

\subsection{Finite Temperature vs. Infinite Temperature}
\label{sec:temperature}

 The majority of the error correction literature assumes an error model akin to an ``infinite temperature limit''.  More precisely, an array of physical qubits receives errors from some set of error operators ${E_i}$ independently at random with some probability $p$ during every error correction cycle.  The threshold theorems state that there exists some critical error probability $p_c$ below which it is possible to return an error correcting code to its encoded state with unit probability for asymptotically large systems; e.g., for the toric code, $p_c\approx.109$ \cite{Wang2009}.
 
 In contrast, thresholds at finite temperature are usually quoted in terms of a critical \emph{temperature}.  That is, there must exist some critical temperature $T_c$ below which codes can be reliably corrected.  Unfortunately, this definition obscures a great deal of physics---different choices of bath model can greatly affect the dynamics of the error processes, to the extent that a quoted ``critical temperature'' often implicitly specifies a choice of bath model.  Because different bath interactions can give rise to different system dynamics, the choice of bath also directly affects the strategy used for error correction.  For example, it is known that the toric code's threshold temperature is altered by considering a space-correlated bath rather than an uncorrelated one \cite{Novais2013}.
 
 The main consequence of choosing an Ohmic bath is that it sets the amplitude of the excitation hopping process.  That is, $\gamma(0)$ is determined by the $\omega \rightarrow 0$ limit of the spectral density of the bath, and for the Ohmic bath taking the $\omega\rightarrow0$ limit of  Eq. \ref{eq:gammadef} yields $\gamma(0) \sim T$. 
Ultimately, this means that the hopping rate of domain walls is controlled by this choice of bath model. At finite temperatures, this introduces correlations into the patterns of errors that effect the system, and so it is no longer possible to talk about an ``independent error probability per site''. 
In contrast to the behavior of $\gamma(0)$, 
the other operationally important feature of the bath, the ratio of defect creation and annihilation rates, is set by detailed balance to Boltzmann-like scaling (i.e., $\gamma(\Delta)/\gamma(-\Delta) = \exp(-\Delta/T)$), and is independent of the choice of bath spectrum.
 
 In the most extreme case, at sufficiently low temperature, pairs of neighboring defects are most often immediately dissipated by the bath upon creation via a $D_b$ operator.  However, if a pair creation is followed by a pair hopping event -- i.e., a $D^\dagger_{b}$ followed by a $T_{b}$ -- the error can no longer be immediately dissipated by the local action of the bath.  Subsequently the defects will undergo a one-dimensional random walk, and topologically non-trivial random walks will cause uncorrectable logical errors.
 
 Thus, error correcting the 1D Ising model at low temperature with this sort of bath dynamics reduces to attempting to identify these randomly-migrating rare pairs of defects.  While a majority-rule decoding scheme works in both low and high temperature limits for the Ising model, if the number of measurement resources is restricted, the standard majority rule scheme breaks down because of the intrinsic uncertainty regarding unmeasured defects.

%%%%%%%%%%%%%%%%%%%%%%
%%%%%%%%%%%%%%%%%%%%%%
\section{Few Measurement Error Correction Algorithm}
\label{sec:ecc_alg}
%%%%%%%%%%%%%%%%%%%%%%

\subsection{The Algorithm}

In this section, we sketch a new algorithm which reliably removes errors in the 1D Ising model below a threshold temperature, which we determine numerically.  The primary technical innovation of this algorithm, and its generalization to quantum memories based on any stabilizer Hamiltonian, is that it does not require measurement of the complete set of stabilizer operators for a given stabilizer code---only a fixed subset.  We assume $i)$ that the system is subject to periodic measurements on periodically spaced measurement ``patches'', $ii)$ that measurement readout and processing occurs much faster than any system timescale, and $iii)$ that the system is subject to a thermal bath as described in Sec. \ref{sec:Defs}.

\begin{figure}
\begin{center}
\scalebox{1}{\includegraphics[width=1.0\columnwidth]{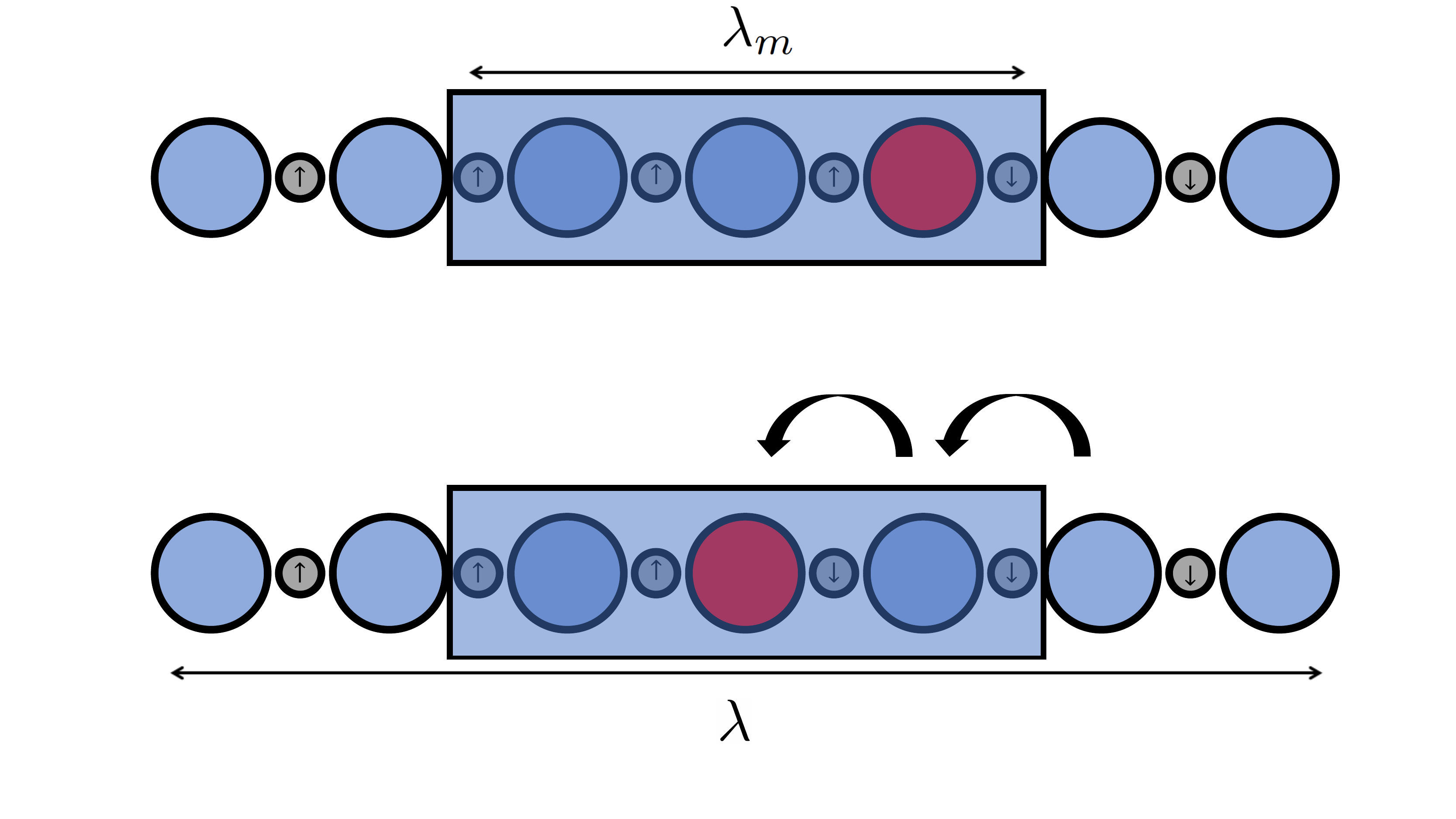}}
\end{center}
\caption{This cartoon illustrates the ``centering'' procedure for detected defects on a unit cell.  Spin variables are in gray, and domain wall variables are in blue (no defect present) and red (defect present).  When a defect is detected on a measurement patch (blue box), it is swapped to the center of the measurement patch via the $\rm{DSWAP}$ operator (black arrows).  The defect immediately adjacent to it is also swapped onto the measurement patch so as not to pull apart defect pairs that would have otherwise dissipated.  Measurement patch lengthscale $\lambda$ indicated by arrow on top, and unit cell lengthscale $\lambda$ indicated by arrow on bottom. The measurement fraction is defined as $m\equiv \lambda_m/\lambda$.}
\label{fig:detect_swap}
\end{figure}

The algorithm can be summarized in five steps:

1. Measure stabilizers on patches, keeping record of the age of defects that are already on patches---i.e., the amount of time a defect is continuously detected on a patch---as well as defect locations.  

2. Perform ``centering'' on patches with defects (see \figref{fig:detect_swap}), based on centering protocol introduced in Ref. \onlinecite{Freeman2017}.

3. Calculate probability of fusion (explicitly given in Eq. \ref{eq:bayes_simple}) for all pairs of measured defects residing on the measurement sites.  This probability serves as an estimate for whether two defects should be paired or not for the purposes of error correction.

4. Probabilistically perform error correction based on probabilities calculated in step 3.

5. Repeat steps $1-4$.

Step 2 encourages defects to remain localized at measurement patches.  This centering protocol can be performed entirely unitarily by the $\rm{DSWAP}$ operator, which takes the following form in the Pauli basis,
\begin{align}
\rm{DSWAP}_i &= \frac{1}{2} \left(III + I\sigma_x I + \sigma_z I \sigma_z - \sigma_z \sigma_x \sigma_z \right)
\end{align}

\noindent where $i$ indexes the location of the first qubit being acted upon by the operator by convention.

If a domain wall exists either between the first and second qubit or the second and third qubit, then the \rm{DSWAP} operator exchanges those domain walls.  If there are no domain walls, it acts as the identity.  By concatenating a sequence of \rm{DSWAP}s, i.e. $\rm{DSWAP}_i \rm{DSWAP}_{i+1} \rm{DSWAP}_{i+2} ...$, domain walls can be shuttled to the center of the measurement patch for efficient tracking.

The centering process, illustrated in Fig. \ref{fig:detect_swap}, aids the probability of fusion calculation by ensuring that the coordinates and measurement times are representative of when and where defects are actually created.  If defects escape from measurement patches, then upon being measured again, the time recorded by the measurement patch now underestimates how old the defect actually is, biasing the probability estimate.  This centering operation greatly reduces the probability of defect escape.  Any remaining underestimate of defect lifetimes can be fixed by a more elaborate record keeping protocol (See Appendix \ref{sec:appendix_fix_long_errors}).

 Note that the pattern of $\rm{DSWAP}s$ used in \figref{fig:detect_swap} also swaps the neighboring, \emph{unmeasured} defect onto the measurement patch.  This is to ensure that the protocol does not inadvertently create a new separated pair of defects in the system by shifting only one defect in a potentially adjacent pair. 
 
\subsection{Fusion Probability Calculation}
\label{sec:bayes_prob_calc}

%************************TRIM SUBSTANTIALLY AND PUT IN APPENDIX

To perform error correction properly, we need to be able to estimate the probability that two given measured defects are a pair, given that they have been measured at two particular measurement patches at two different times.  For notational convenience, we define:
\begin{align}
d_1 : d_2 \equiv {\rm defect}\ d_1\ {\rm and}\ d_2\ {\rm are\ a\ pair}
\end{align}
and
\begin{align}
d_i^{x_1,t_1} \equiv {\rm defect}\ d_i\ {\rm measured\ at\ time}\ t_1\ {\rm at\ patch}\ x_1
\end{align}

\noindent Then, we aim to calculate the fusion probaility:
\begin{align}
P(d_1:d_2 | d_1^{x_1,t_1} \wedge d_2^{x_2,t_2}),
\end{align}
i.e., the probability that two detects measured at spacetime coordinates $(x_1,t_1)$ and $(x_2,t_2)$ are part of the same defect pair, and therefore should be fused in a correction step.

To calculate this probability, we proceed via Bayes rule: 
\begin{align}
\label{eq:bayes}
&P(d_1:d_2 | d_1^{x_1,t_1} \wedge d_2^{x_2,t_2}) = \\ \notag
&\frac{P(d_1^{x_1,t_1} \wedge d_2^{x_2,t_2} | d_1:d_2) P(d_1 : d_2)}{P(d_1^{x_1,t_1}\wedge d_2^{x_2,t_2})}
\end{align}
The individual terms on the right hand side of equation \ref{eq:bayes} are straightforward to interpret.  $d_i^{x_i,t_i}$ indicates a defect  residing on a measurement patch centered on spacetime coordinate $x_i,t_i$.  $P(d_1^{x_1,t_1} \wedge d_2^{x_2,t_2} | d_1:d_2)$ represents the probability that two measured defects would be at $(x_1,t_1)$ and $(x_2,t_2)$ given that they are indeed a pair. $P(d_1 : d_2)$ represents the probability that two measured defects, $d_1$ and $d_2$, are in fact a pair.  Finally, $P(d_1^{x_1,t_1}\wedge d_2^{x_2,t_2})$ is the probability that two defects are measured, one at $(x_1,t_1)$, and the other at $(x_2,t_2)$.

$P(d_1^{x_1,t_1} \wedge d_2^{x_2,t_2} | d_1:d_2)$ can be related to the probability that a one dimensional diffusion process with diffusion constant $D$ will perform an excursion with a displacement $|x_2-x_1|$ or greater in a time $t_2 - t_1$, \textit{i.e.,} will perform an excursion that can reach measurement patches at $x_1$ and $x_2$. Explicitly,
\begin{align}
\label{eq:bayesformula}
&P(d_1^{x_1,t_1} \wedge d_2^{x_2,t_2} | d_1:d_2) \\ \notag
&= 1-2\int_{0}^{|x_2-x_1|} dx \frac{1}{2\pi D |t_2-t_1|} {\rm exp}\bigg(-\frac{|x_2-x_1|^2}{2 D |t_2-t_1|}\bigg) \\ \notag 
&= 1 - {\rm erf}\bigg(\frac{|x_2-x_1|}{2\sqrt{D |t_2 - t_1|}}\bigg) 
\end{align}
For our analysis, we will choose $D\propto \gamma_0$.  The exact correspondence between $D$ and $\gamma_0$ depends on the details of the error correction algorithm itself, so, in practice, we treat the constant of proportionality as an empirically tuned parameter.  Furthermore, we approximate any detected defects as arising from a pair that was created an equal distance between the measurement patches at locations $x_1$ and $x_2$ for the purposes of calculating the probability in Eq. \ref{eq:bayesformula}.

 As we discuss in Appendix \ref{sec:bayesdecoder}, the remaining two factors are not as important for the decoding scheme as the likelihood term in Eq. \ref{eq:bayesformula}.  In practice, we find that using the expression from Eq. \ref{eq:bayesformula} alone is sufficient to provide resilient error correction.  We defer further discussion to the appendix.

 \subsection{Error dynamics}
  
  In this section, we discuss parameter regime in which we expect the error correcting algorithm to perform well.  We then derive the logical error rate for a simple error model, under some simplifying assumptions about the error dynamics.  While this error model does not account for the complete error dynamics of the full 1D Ising model in the presence of our protocol, we argue how it nonetheless serves as a worst-case approximation to the true error dynamics.  Finally, we describe how this protocol provides a threshold for any finite density of measurements.
 
 \subsubsection{Correspondence between model and full error dynamics}
 \label{sec:correspondence}
 
 In this section, we detail the approximations and rate assumptions that are necessary for the error correcting algorithm to perform well.  The primary approximations made are concerning \textbf{(1)} fast defect detection, \textbf{(2)} accurate pairing, \textbf{(3)} defects escaping measurement patches, and \textbf{(4)} defect interactions.
 
 \textbf{(1)}. If defects are produced in between measurement patches faster than they are detected, then this algorithm cannot in principle correct errors.  Thus, we require that the characteristic diffusion time for defects in the bulk to migrate to a measurement patch, $\gamma_0^{-1}(\lambda_b)^2$, where $\lambda_b=\lambda-\lambda_m$, to be much shorter than the characteristic timescale over which a pair of defects is created in the bulk, $\gamma_0^{-1} \exp{\Delta/T}$.  Thus, working at low temperature ensures the validity of this approximation.
 
 \textbf{(2)}. If defects are paired incorrectly more often than they are paired correctly, then the algorithm will fail.  Let $\tau_{\epsilon}^{-1}$ be the rate of the error process and $\tau_d^{-1}$ be the rate of a non-erroneous error correction operation.  To ensure that $(\tau_{\epsilon}^{-1} / \tau_d^{-1})\leq1$ (see Eq. \ref{eq:err_scaling}), we must work in the diffuse limit, where the average number of defects per unit cell is much less than $1$.  This is equivalent to $\lambda \gamma_+<<1$.  This ensures that, when defects are being processed by the algorithm, that more often than not, defects will be correctly paired simply because it's unlikely there are any other defects nearby.  Thus, assuming condition (1)---that defects are detected quickly---defect pairs satisfying Eq. \ref{eq:bayesformula} are more likely than not to be genuine pairs. 
 
 \textbf{(3)}. While the simple model does not account for defects escaping measurement patches, this can occur in the real system when a series of translation events occurs between measurements.  For a measurement rate $\chi$, these processes are of $O(\gamma_+ (\gamma_0/\gamma_-) (\gamma_0/\chi)^{\lambda_m/2})$, for measurement patches of size $\lambda_m$, assuming $\chi > \gamma_0$.  Thus, this process can be suppressed by working with a larger sized measurement patch, or with a measurement rate suitably larger than the intrinsic translation rate of the system, $\gamma_0$.  At worst, the age of defects that escape measurement patches but that are subsequently recaptured may be underestimated by the algorithm, because the ``age'' of the defect would be erroneously reset to zero.  This would then erroneously underestimate the distance the algorithm would plausibly search for a pairing defect---i..e, the denominator of the error function in Eq. \ref{eq:bayesformula} could be artificially small because the ``real'' defect age is actually older.  While these sorts of errors can potentially spoil the error correcting protocol at very long distances---much larger than considered in this manuscript---these errors can be corrected with a modified version of our algorithm without any additional measurement resources, detailed in Appendix \ref{sec:appendix_fix_long_errors}.

 \textbf{(4)}. In reality, defects can annihilate without the protocol intentionally pairing them.  To leading order, at low temperature, these processes are ``self-correcting''.  That is, a pair of neighboring defects enters the system, and then is subsequently annihilated.  In principle, it is possible for a sequence of $k$ free pairs of defects to appear in the bulk---one pair per unit cell---of which, $k-1$ are then subsequently erroneously ``corrected'', resulting in two defects separated by a distance $k \lambda$, but this process is exponentially slow in the average defect unit cell density, which we already choose to be small via condition \textbf{(2)}.  That is, the error rate due to the erroneous  separation of defects by a distance $\lambda k$ is $\propto(\lambda \gamma_+)^k$.

 \subsubsection{A simple error model}
 \label{sec:threshold}
 
 To bound the error rate of the Ising model in the presence of our protocol, we study a simple error model for ``spurious error correction'' events.  A representative example of one of these events is when two pairs of defects are detected in the system (four defects total on four distinct measurement patches), and the protocol erroneously pairs one defect from each distinct pair.  Because the density of defects is low at low temperature, this error process is similar to an error process that occasionally randomly translates one defect of a pair some distance.  The distance one of the pair becomes separated depends on the age of the defect, as well as whether the erroneously paired defect was to the left or the right of the original pair of defects.
 
 Thus, the simplified error model is defined as follows:  suppose that a single pair of defects is in the system, and that no new pairs will be introduced.  One of the pair is fixed on a measurement patch, and the other is, at time $t=0$, undetected and residing somewhere in the bulk between measurement patches.  We will model spurious error correction by an error process that translates the unmeasured defect by a distance $2\sqrt{\gamma_0 \delta t}$. As time increases, the characteristic distance over which this error process can occur also increases, in accordance with the typical pair-wise separation between two defects performing a random walk.  This typical distance is exactly the factor used by the error correction algorithm to determine if a pair of defects should be corrected or not.
 
 An uncorrectable error will occur if the bulk defect remains undetected up until it crosses half the system.  For unit cells of size $\lambda$, the probability that this occurs is roughly $(\tau_{\epsilon}^{-1} / \tau_d^{-1})^{k}$, where $k$ is the number of times the error process must occur for the error process to have separated the defects a distance $(L/2\lambda)$, $\tau_{\epsilon}^{-1}$ is the rate of the error process and $\tau_d^{-1}$ is the rate of a non-erroneous error correction operation.  After a timescale $q\cdot\tau_{d}$, the defect pairs will have been separated a distance equal to

\begin{equation}  
\sqrt{q \cdot \gamma_0 \tau_{d}}
\end{equation}
 assuming that they are never correctly paired.  This grows as $q^{1/2}$, thus, $k$ scales approximately as $(L/\lambda)^{2}$.
Finally, assuming there are $L/\lambda$ such simultaneous independent error processes in the system---one for each measurement patch---then the total error probability scales as
\begin{align}
\label{eq:err_scaling}
 P(\rm{error})\leq (L/\lambda) (\tau_{\epsilon}^{-1} / \tau_d^{-1})^{(L/\lambda)^{2}}.
\end{align}
For sufficiently low temperatures (see Sec. \ref{sec:correspondence}), $\tau_{\epsilon}^{-1}$ is much smaller than $\tau_d^{-1}$, thus the full probability of erroneous corrective operations is exponentially small in system size.

While our toy error model is ``non-interacting''---that is, it assumes $L/\lambda$ independent error processes which, in sum, take the form described in Eq. \ref{eq:err_scaling}---a more careful treatment of the error process, including interactions between defects, as in the real model, would result in an error probability \emph{smaller} than the one calculated here.  In Sec. \ref{sec:experiments}, we provide numerical evidence that the lifetime of the Ising model in the presence of the protocol scales exponentially with the number of measurement patches, as anticipated by the upper bound in Eq.\ref{eq:err_scaling}.

 \subsubsection{Error correction at any measurement density}
 \label{sec:ecc_any_m}
 
 A key feature of the protocol is the ability to provide an error correcting threshold temperature at any finite measurement density.  In particular, for a fixed measurement density $m$, and fixed measurement and bulk length scales $\lambda_m$ and $\lambda_b$, respectively, it is still possible to satsify the rate assumptions of Sec. \ref{sec:correspondence} by tuning temperature sufficiently low.  Each rate assumption does not explicitly depend on total system size $L$, only unit cell size $\lambda$.
 
 In practice, larger $\lambda_b$ (alternatively, smaller $m$) will result in lower threshold temperatures simply because the temperature must be lower to satisfy the rate assumptions of conditions \textbf{(1)} and \textbf{(2)}.  We provide explicit evidence of this scaling in Fig. \ref{fig:CriticalTempVsMeasurementFraction}.

 %%%%%%%%%%%%%%%%%%%%%%
%%%%%%%%%%%%%%%%%%%%%%
\section{Finite Temperature Simulations}
\label{sec:experiments}
%%%%%%%%%%%%%%%%%%%%%%

In this section, we present the numerical simulations of the protocol on finite-size systems of length $L$. We consider systems with unit cells of size $\lambda$ with $\lambda_m=3$ measured sites in each unit cell, and a measurement fraction of $m\equiv \lambda_m/\lambda$.

\begin{figure}
\begin{center}
\scalebox{1}{\includegraphics[width=1.0\columnwidth]{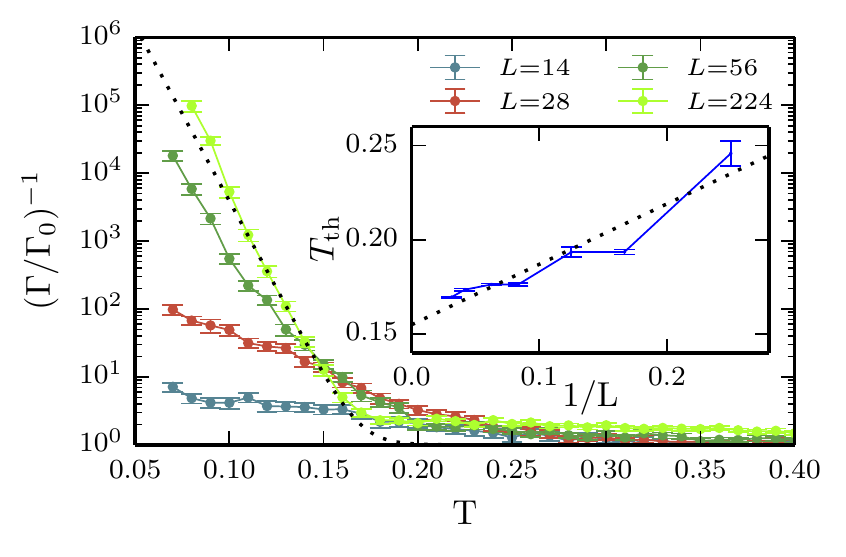}}
\end{center}
\caption{Lifetime {\it enhancement} of several system sizes $L$ as a function of temperature $T$ using a measurement fraction of $m=3/7$.  Dotted line indicates example fit for threshold temperature $T_{th}$ extraction for $L=224$ data.  Below $T\approx0.16$, we find the lifetime to grow exponentially with $L$, indicative of a finite-temperature threshold. {\it Inset}: finite size scaling of $T_{th}$ with inverse length $1/L$.  Extrapolation to the infinite system limit yields a threshold temperature of $T_{\rm{th}}=.155(6)$.}
\label{fig:LifetimeVsTemperature}
\end{figure}

\figref{fig:LifetimeVsTemperature} depicts the scaling of the system lifetime enhancement with temperature for several system sizes.  Below a certain temperature, the system lifetime increases exponentially with system size.  Due to finite size effects, it is difficult to extract an unambiguous threshold temperature, but below $T\approx0.16$, the lifetime increases exponentially with larger system size.  We estimate the threshold by fitting $\Gamma(T)^{-1}$ to $1+\rm{exp}(-a*(T-T_{\rm{th}}))$. The inset of \figref{fig:LifetimeVsTemperature} shows the finite-size scaling of $T_{\rm{th}}$, which suggests that $T_{\rm{th}}$ remains non-zero in the limit of $L\rightarrow \infty$; this demonstrates that this protocol has a finite-temperature threshold in the thermodynamic limit.  In this limit, we find $T_{\rm{th}}=.155(6)$.

%%%%%%%%%%%%%%%%%%%%%%
\begin{figure}
\begin{center}
%\subfloat{\scalebox{1}{\includegraphics[width=\columnwidth]{LifetimeVsSystemSize.pdf}}}
%\\
%\vspace{-1.2\baselineskip}
%\subfloat{\scalebox{1}{\includegraphics[width=\columnwidth]{LifetimeVsSystemSizeReduced.pdf}}} 
\scalebox{1}{\includegraphics[width=1.0\columnwidth]{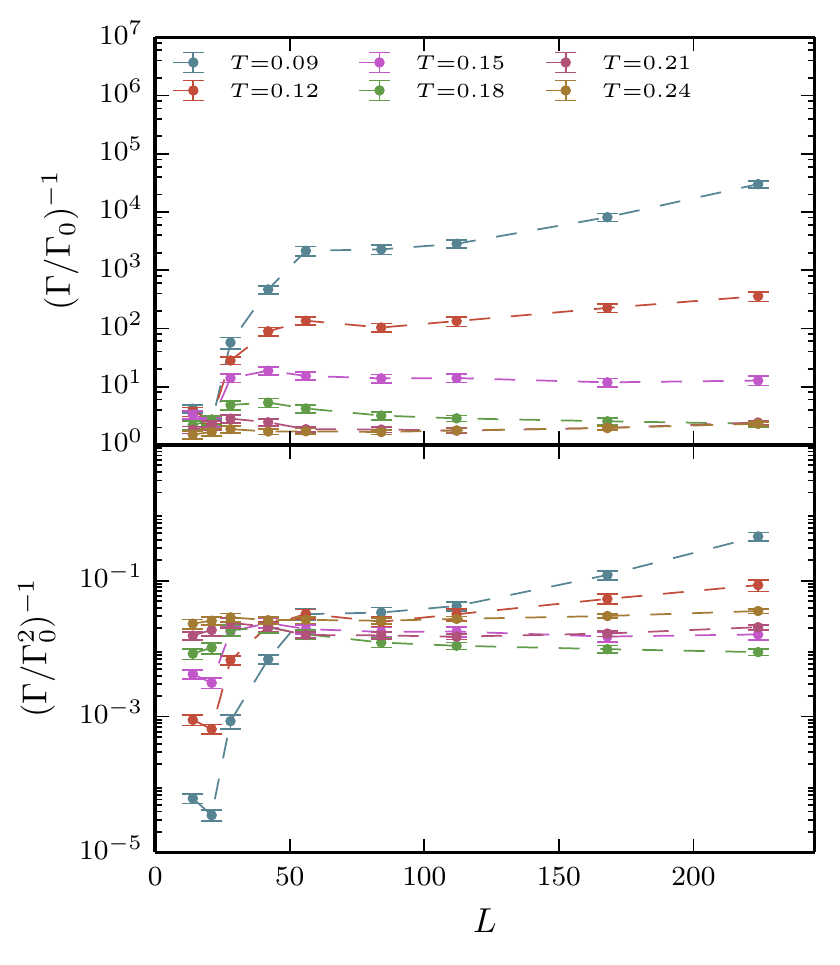}}
\end{center}
\caption[justification=raggedright]{
(Top) Lifetime enhancement for temperatures both below and above the threshold as a function of system size $L$, using a measurement fraction $m=3/7$.  Note the monotonic growth in lifetime at low temperatures, as well as the plateau in lifetime for moderately sized systems. A $3/7$ measurement fraction was used for this data.
(Bottom) The same data rescaled by an additional factor $\Gamma_0^{-1}$, to emphasize the origin and scaling of the plateau in system lifetimes.}
\label{fig:LifetimeVsSystemSize}
\end{figure}
%%%%%%%%%%%%%%%%%%%%%%

\figref{fig:LifetimeVsSystemSize} (top) depicts the finite-size scaling of the lifetime enhancement for temperatures below and above the threshold.  Note that for systems above $T\approx0.16$, larger system sizes asymptote to a constant lifetime enhancement, whereas for models below $T\approx0.16$, the lifetime grows monotonically with system size.  We find that beyond $L=100$, finite size effects are significantly reduced, as small-system sizes cannot easily suppress second order errors, such as defects escaping from measurement patches or multiple pairs of defects in the system.  Such errors are actually uncorrectable for systems where $L/\lambda \leq 4$---hence the plateau appearing around $L=50$ to $L=100$.  Above these system sizes, the exponential scaling returns.  This plateau is of height $\mathcal{O}(\Gamma_0^{-2})$---the characteristic timescale of these ``second-order'' events.  This scaling is made apparent in \figref{fig:LifetimeVsSystemSize} (bottom), where the lifetime has been scaled by of $\Gamma_0^{-2}$ for each temperature.

%************************PUT IN TERMS OF SYSTEM SIZE
%************************MAYBE REMOVE INTERMEDIATE SYSTEM SIZES - remove 12,16,24 from Fig. 2 or maybe just 4,8,16,32.
%************************Mush together figure 3 and 4 maybe?
%************************Lines in fig. 5 and fig. 7 to guide the eye
%************************include m and system sizes in figure captions
%************************maybe have a plot for Delta critical temperature scaling
%************************make the toric code figure sparser perhaps?

\begin{figure}
\begin{center}
\scalebox{1}{\includegraphics[width=1.0\columnwidth]{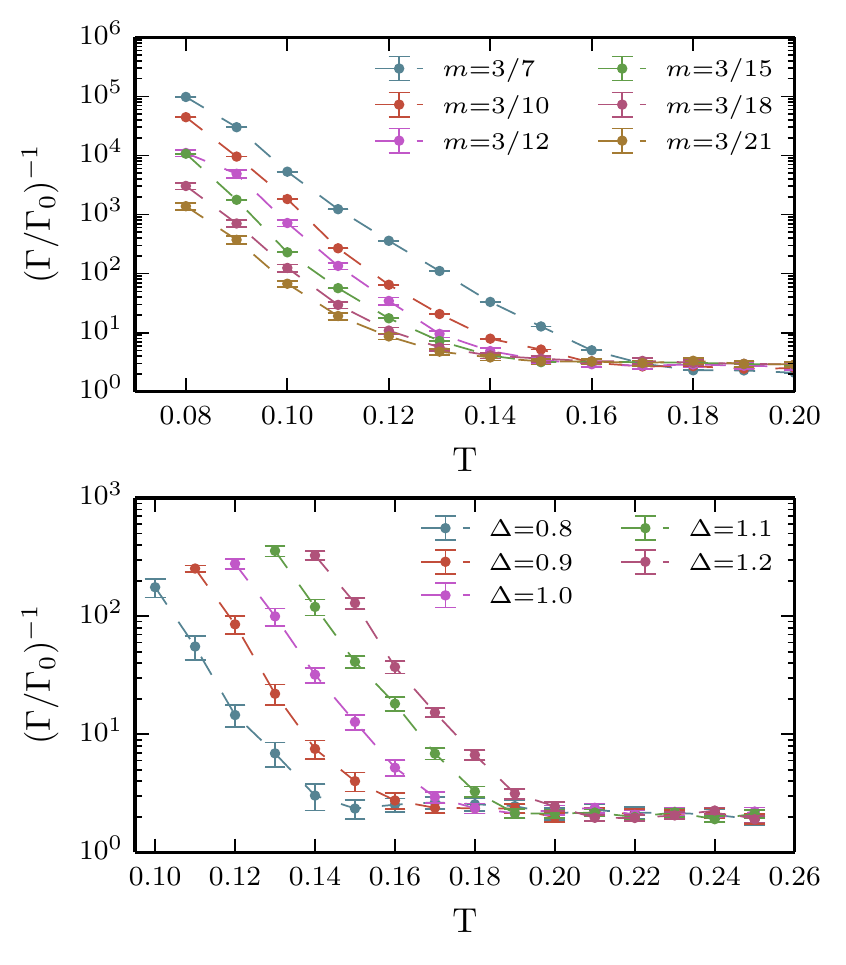}}
\end{center}
\caption[justification=raggedright]{
(Top) The lifetime enhancement as a function of temperature for several measurement fractions $m$ of $32$ (i.e., $L=32\cdot 3/m$.  Note the threshold temperature decreases  with $m$.
(Bottom) Lifetime enhancement as a function of temperature for a variety of energy scales $\Delta$ (see Eq. \ref{eq:isingham}) for an $L=224$ system with m=$3/7$.}
\label{fig:LifetimeVsMeasurementFractionVsDelta}
\end{figure}

\begin{figure}
\begin{center}
%\subfloat{\scalebox{1}{\includegraphics[width=\columnwidth]{CriticalTempVsMeasurementFraction.pdf}}}
%\\
%\vspace{-1.2\baselineskip}
%\subfloat{\scalebox{1}{\includegraphics[width=\columnwidth]{CriticalTempVsDelta.pdf}}}
\scalebox{1}{\includegraphics[width=1.0\columnwidth]{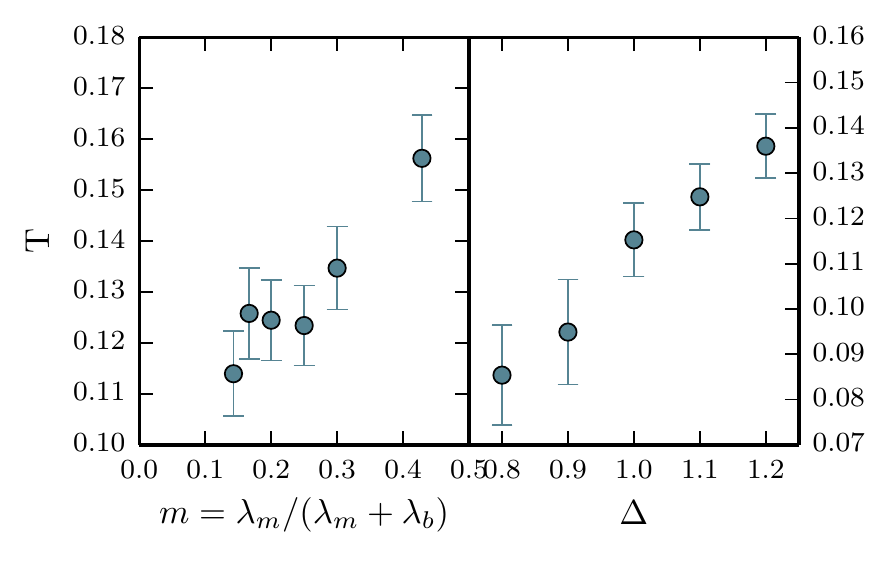}}
\end{center}
\caption[justification=raggedright]{The threshold temperatures as a function of measurement fraction, $m$ (left),  and the Hamiltonian energy scale $\Delta$ (see Eq. \ref{eq:isingham})(right).  At zero measurement fraction, the critical temperature is 0, and at unit measurement fraction, the critical temperature is $\mathcal{O}(1)$.}
\label{fig:CriticalTempVsMeasurementFraction}
\end{figure}

\figref{fig:LifetimeVsMeasurementFractionVsDelta} depicts the lifetime enhancement as a function of temperature for several different measurement fractions as well as different energy scales, $\Delta$.  It is evident that measuring a smaller fraction of the lattice causes the threshold temperature to shift downwards.  This dependence of the threshold temperature on the measurement fraction is depicted explicitly in \figref{fig:CriticalTempVsMeasurementFraction} (left panel).

%%%%%%%%%%%%%%%%%%%%%%
%%%%%%%%%%%%%%%%%%%%%%

%************************EXPAND THIS PARAGRAPH A LITTLE

The scaling of the threshold temperature with $\Delta$ is presented in  \figref{fig:CriticalTempVsMeasurementFraction} (right panel). By contrasting the left and right panels of Figure \ref{fig:LifetimeVsMeasurementFractionVsDelta}, one can deduce the relative benefits of error suppression via more measurement resources versus error suppression via hamiltonian engineering (i.e., a larger gap to excitation).  %In particular, it's interesting that while the lifetime increases linearly with energy gap, $\Delta$, (at least for the energy scales we have simulated) the increase in lifetime with measurement fraction is manifestly non-linear, suggesting a non-trivial relationship between entropy reduction and memory lifetime.

  %%%%%%%%%%%%%%%%%%%%%%
%%%%%%%%%%%%%%%%%%%%%%
\section{Generalization to Higher Dimension}
\label{sec:tc_alg}
%%%%%%%%%%%%%%%%%%%%%%

 In this section, we sketch how the algorithm presented in Sec. \ref{sec:ecc_alg} generalizes to a higher dimensional stabilizer quantum memory---the $2D$ toric code.  Where the dynamics of the 1D Ising model are typified by one dimensional random walks of defects, the nonequilibrium dynamics of the toric code are driven by two dimensional random walks of quasiparticle excitations.  Consider the toric code hamiltonian:.
 \begin{align}
\Htc &= -\Delta_e \sum_v A_v -\Delta_m \sum_p B_p ,\label{eq:HTC}\\
A_v &\equiv \prod_{j \in v} \sigma_j^z,\quad B_p \equiv \prod_{j \in p} \sigma_j^x,\label{eq:AvBp}
\end{align}
\noindent
where $v$ denotes the 4-spin vertices of the square lattice, and where and $p$ denotes the 4-qubit plaquettes on the edges of a 2D square lattice\cite{Kitaev2003}.
While domain wall excitations in the 1D ising model are associated with -1 eigenstates of the $\sigma_z^i \sigma_z^{i+1}$ stabilizers, quasiparticle excitations for the toric code are associated with -1 eigenstates of the $A_v$ and $B_p$ stabilizers as defined in Eq. \ref{eq:AvBp}.
 
\begin{figure}
\begin{center}
\scalebox{1}{\includegraphics[width=1.0\columnwidth]{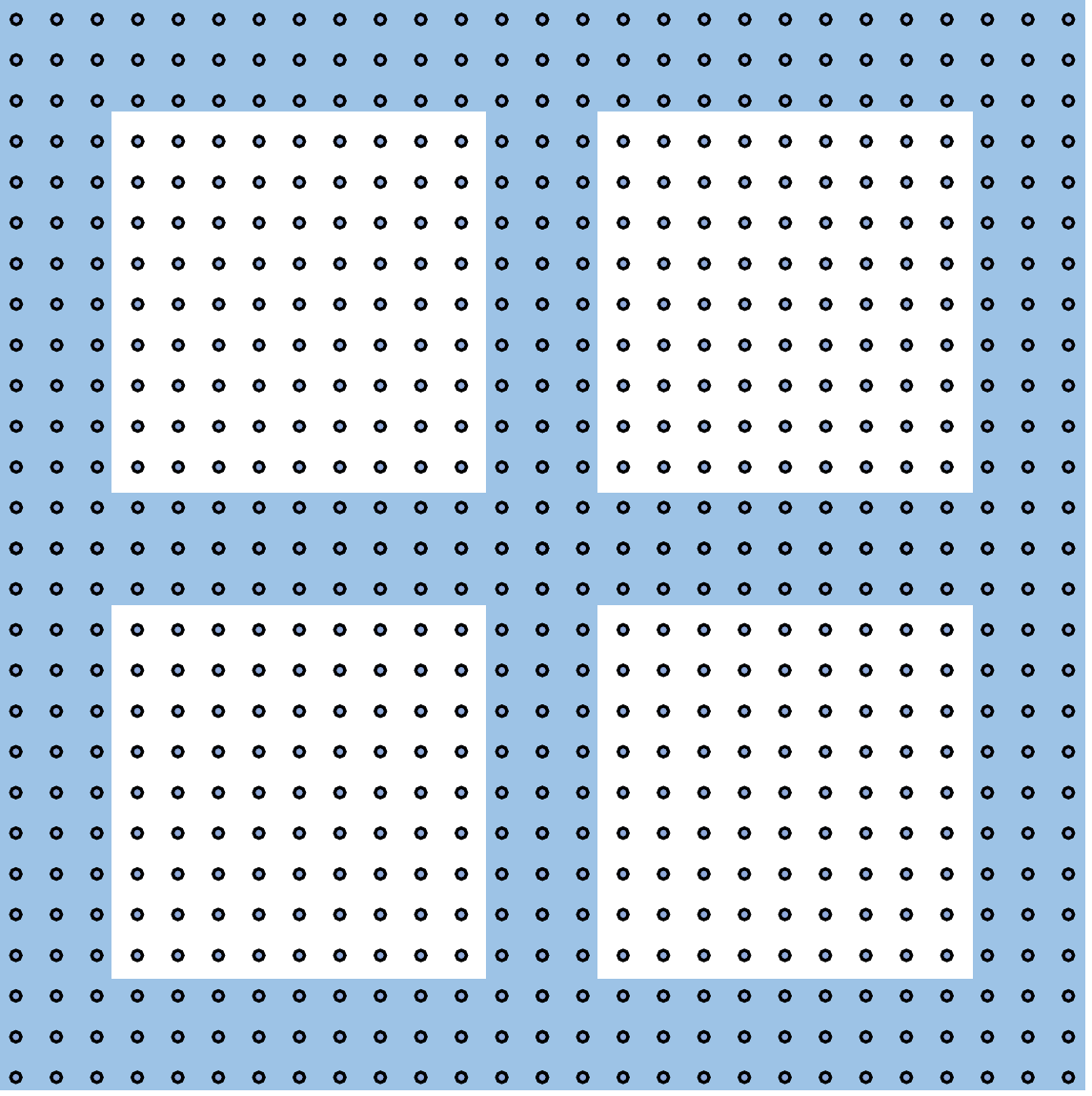}}
\end{center}
\caption{Here we sketch one possible $m=63/144$ (for a 12x12 unit cell) geometry for the measurement rails for a realization of our protocol.  More sparse geometries can be realized simply by moving the rails of measurement farther apart.  Measured sites are in light blue, and vertex locations for the toric code are circles.  Spin variables (not pictured) reside directly between neighboring vertices.}
\label{fig:tc_fig}
\end{figure}

 Broadly speaking, the algorithm is identical, but instead of having ``patches'' of measurement, there are measurement ``rails'', as indicated in \figref{fig:tc_fig}.  One such set of rails must exist for both types of excitations in the toric code---that is, one for the $B_p$ stabilizers, and another set of measurement rails for the $A_v$ stabilizers.  The error detection and correction can then be performed completely in parallel for both types of excitations, as they are independent.  ``Centering'' of defects on rails amounts to shift-swapping defects into the center of the measurement rail\cite{Freeman2017}.  We conjecture that a sparse measurement strategy with randomly placed measurement patches of fixed diameter might exist for sufficiently large and sufficiently cold systems.  However, the rail geometry of \figref{fig:tc_fig} is the simplest geometry that allows us to argue for a threshold temperature for the toric code, based on a generalization of the simple error model used for the Ising model in Sec. \ref{sec:threshold}.  The only difference between the upper bound to the expected scaling of the probability of uncorrectable errors in the toric code versus the expected scaling of the Ising model (i.e., Eq. \ref{eq:err_scaling}) is the prefactor becomes $\propto (L/\lambda)^2$ for the toric code instead of $L/\lambda$, where $L$ represents the linear dimension of the toric code.

 %%%%%%%%%%%%%%%%%%%%%%
%%%%%%%%%%%%%%%%%%%%%%
\section{Discussion}
\label{sec:discussion}
%%%%%%%%%%%%%%%%%%%%%%

We have provided numerical and theoretical evidence of a limited measurement error correction protocol for a stabilizer code with string-like excitations.  The primary technical innovation of our algorithm is a Bayesian decoding scheme for pairing defects based on partial information, sketched in Sec. \ref{sec:ecc_alg}. When combined with the measurement-free defect localization technique developed in Ref. \onlinecite{Freeman2017}, this decoding scheme performs error correction efficiently and results in a stable quantum memory at temperatures below an empirically determined threshold temperature.  So long as an appropriate geometry of measurement devices is in place, and so long as defects undergo diffusive motion via coupling to a thermal bath, this scheme can be extended to higher dimensional stabilizer systems like the toric code, as demonstrated in Sec. \ref{sec:tc_alg}.

Our results for variable measurement fraction complement what is known about decoders in the presence of \emph{noisy} measurements\cite{Nickerson2016}.  Figures \ref{fig:LifetimeVsMeasurementFractionVsDelta} (top) and \ref{fig:CriticalTempVsMeasurementFraction} (left) demonstrate how a reduction in the measurement fraction in the lattice corresponds to a concomitant decrease in the threshold temperature, similar to how thresholds are known to be reduced when increasing the noise on measurements.

More fundamentally, our algorithm can be understood as an entropy reduction scheme.  Configurations that give rise to errors in the encoded subspace are exponentially suppressed as system size is made larger.  This is in contrast with ``energetic'' suppression---that is, suppression by widening the gap to excitations, $\Delta$ (or equivalently lowering the operating temperature).  This tradeoff between entropic and energetic contributions is depicted in Figures \ref{fig:LifetimeVsMeasurementFractionVsDelta} (top versus bottom) and \ref{fig:CriticalTempVsMeasurementFraction} (left versus right).  Depending on the resource requirements of a particular architecture, the threshold temperature can be tuned either by engineering a larger gap, $\Delta$, or by changing the number of measurements used.  In practice, this will depend on the lowest effective temperature available, the maximum measurement rate, as well as the practical difficulty of employing more measuring devices.

%%%%%%%%%%%%%%%%%%%%%%
\section{Acknowledgments}
This material is based upon work supported by DARPA under Grant No. 3854-UCB-AFOSR-0041 and by the National Science Foundation under Grants No. PIF-0803429 and No. CHE-1213141.  CDF was supported by the NSF Graduate Research Fellowship under Grant DGE-1106400 and by the DOE Office of Science Graduate Student Research (SCGSR) program under contract number DESC0014664. MS was supported by the Laboratory Directed Research and Development program at Sandia National Laboratories.  Sandia National Laboratories is a multimission laboratory managed and operated by National Technology and Engineering Solutions of Sandia, LLC., a wholly owned subsidiary of Honeywell International, Inc., for the U.S. Department of Energy's National Nuclear Security Administration under contract DE-NA-0003525.

%%%%%%%%%%%%%%%%%%%%%%

\appendix
%%%%%%%%%%%%%%%%%%%%%%
\section{Bayesian Decoding for the Ising Model}
\label{sec:bayesdecoder}
%%%%%%%%%%%%%%%%%%%%%%

In this section, we provide further discussion of Eq. \ref{eq:bayes}, as well as analytic and numerical arguments for how it can be more simply approximated.

First, we decompose $P(d_1 : d_2)$ into two pieces: a combinatorial piece, and a dynamical piece.

For the combinatorial piece, note that a necessary condition for pairing to be possible is for both defects belonging to a pair to actually be measured.  That is, there might be a large number of measured defects ${d_1, d_2, ..., d_{n_m}}$, but the pairing defect for some of these defects might not be measured. Among those defects which are both measured, and which have their pair also measured, then the probability of selecting two defects that are a pair is simply the combinatorial factor $1 / \binom{N_{\rm{measured\ pair\ defects}}}{2}$ where $N_{\rm{measured\ pair\ defects}}$ counts the average number of measured defects for which their pair is also measured.

The dynamical piece is the probability that $d_1$ and $d_2$ are defects whose pairs are also measured.  This probability depends on how quickly defects make excursions to measurement sites, as well as how quickly defects are being paired---either erroneously or correctly---by the protocol.  We can crudely lower bound this by taking the equilibrium defect distribution, and calculating the probability that a pair of defects lands on a measurement patch.  $\lambda_m / \lambda$ sites have measurement operators, thus $(\lambda_m / \lambda) L \gamma_+$ is an underestimate of the number of defects on measurement patches.  This is an underestimate because the protocol is actually more efficient at concentrating defects on measurement patches than equilibrium dynamics is.  Given $L \gamma_+$ pairs, this amounts to a binomial counting argument, and the expected number of measured pairs is simply $L \gamma_+ (\lambda_m / \lambda)^2$.  Thus, a lower bound to the equilibrium probability of two selected defects being a measured pair is simply $(L \gamma_+(\lambda_m / \lambda)^2) / (L \gamma_+) = (\lambda_m / \lambda)^2$.

As mentioned in Sec. \ref{sec:bayes_prob_calc}, $P(d_1^{x_1,t_1}\wedge d_2^{x_2,t_2})$ is the probability that two defects are measured, one at $(x_1,t_1)$, and the other at $(x_2,t_2)$.  This can be decomposed:

\begin{align}
&P(d_1^{x_1,t_1}\wedge d_2^{x_2,t_2}) = \\ \notag
&P(d_1^{x_1,t_1} \wedge d_2^{x_2,t_2} | d_1:d_2) P(d_1 : d_2) + \\ \notag 
&P(d_1^{x_1,t_1} \wedge d_2^{x_2,t_2} | d_1\cancel{:}d_2) (1 - P(d_1 : d_2))
\end{align}

\noindent The first term is precisely Eq. \ref{eq:bayesformula}, multiplied by $P(d_1 : d_2)$, which we estimated above.  The second term represents the probability of two measurement events, conditioned on those events \emph{not} being part of a pair.  This is essentially the probability that two independent measurement events have occurred, which is approximately the probability that two independent creation events have occurred (assuming defects are measured suitably efficiently).  For a suitably large system at moderately low temperature, this probability can be estimated as $\propto (|t_2-t_1|^2)/((L \gamma_+)^{-2}):=\delta(L,T,\Delta t)$.

In practice, at low temperature for moderately sized systems, $P(d_1 : d_2)$ is very nearly $1$.  This arises from the low density of defects meaning that only rarely are there even a pair of defects in the system.  Of course, if system size is made sufficiently large, this bare probability will become diminished, but it is still the case that defects within a separation distance $2\sqrt{D |t_2 - t_1|}$ are, more often than not, a pair at low temperature.  For the same reason, $P(d_1^{x_1,t_1} \wedge d_2^{x_2,t_2} | d_1\cancel{:}d_2)$ is very nearly $0$ because this probability is roughly equivalent to the probability that two independent pair creation events have occurred, which is unlikely at low temperature and moderate system size.  Again, for sufficiently large systems this probability grows, but it is likewise the case that this probability is small for defects within a distance $2\sqrt{D |t_2 - t_1|}$.  Then, if we write $P(d_1 : d_2) = 1 - \epsilon(L,T)$, and perform some rearranging:

\begin{align}
\label{eq:bayes_simple}
&P(d_1:d_2 | d_1^{x_1,t_1} \wedge d_2^{x_2,t_2}) = \\ \notag
&\frac{1}{1+\frac{P(d_1^{x_1,t_1} \wedge d_2^{x_2,t_2} | d_1\cancel{:}d_2) \epsilon(L,T)}{P(d_1^{x_1,t_1} \wedge d_2^{x_2,t_2} | d_1:d_2) (1-\epsilon(T))}} \\ \notag
\geq &\frac{1}{1+\frac{\frac{\delta(L,T,\Delta t)}{1-\epsilon(L,T)}} {P(d_1^{x_1,t_1} \wedge d_2^{x_2,t_2} | d_1:d_2)}} \\ \notag
\approx &\frac{1}{1+\frac{\delta(L,T,\Delta t)}{P(d_1^{x_1,t_1} \wedge d_2^{x_2,t_2} | d_1:d_2)}}
\end{align}

\noindent Thus, only when $P(d_1^{x_1,t_1} \wedge d_2^{x_2,t_2} | d_1:d_2) << \delta(L,T,\Delta t)$ is this factor not equal to $1$.  This naturally occurs when comparing defects that are much farther apart than diffusive motion would usually allow.  For example: for a very large system, if one defect of a pair is measured at site $0$ and another defect belonging to another independent pair is measured at site $L/2$ shortly thereafter (compared to the timescale for defect motion), it is exceedingly unlikely for these two measured defects to be a pair because it's exponentially unlikely for such a long random excursion to occur.  In this way, the factor $P(d_1^{x_1,t_1} \wedge d_2^{x_2,t_2} | d_1:d_2)$  serves as an indicator function which answers the question, ``Could these two defects have arisen from a random walk starting in the same place?''.  The factor $\delta(L,T,\Delta t)$ sets the cutoff for a plausible excursion---i.e., when the error function is much less than this term, the denominator of \ref{eq:bayes_simple} blows up, and the probability of performing that fusion is essentially zero.

In practice, the precise details of these additional factors arising from Bayes theorem aren't too important for the protocol to function, and we find that using the conditional probability $P(d_1^{x_1,t_1} \wedge d_2^{x_2,t_2} | d_1:d_2)$ itself as a proxy for the full expression from Bayes theorem is sufficient to reliably correct errors.  We provide some heuristic comparisons of different decoding schemes in Appendix \ref{sec:appendix_alt_decode_scheme}.

%%%%%%%%%%%%%%%%%%%%%%
\section{Alternative Algorithm for Estimating Defect Lifetimes}
\label{sec:appendix_fix_long_errors}
%%%%%%%%%%%%%%%%%%%%%%

The algorithm, as presented in Sec. \ref{sec:ecc_alg}, is susceptible to errors due to systematically underestimating defect lifetimes.  In practice, this error rate is small---small enough that it was not detectible in our numerical studies---but it is nonetheless present and has the potential to spoil the increase in lifetime with system size for the protocol.  In this section, we outline how this problem introduces a system size independent uncorrectable lengthscale into the algorithm, and we provide an alteration to our presented algorithm that can account for these errors, restoring the expected system size scaling.

\subsection{The maximum correctible lengthscale}

 If we denote the timescale over which defects escapes measurement patches on average as $\te$, then a defect pair that is separated by much more than $\sqrt{D \te}$ will be overwhelmingly likely to escape from its measurement patch before the denominator of Eq. \eqref{eq:bayesformula} can grow large enough to match the defect to its pair---potentially spoiling the system size scaling of the algorithm.
 
 In practice, by occasionally scaling the size of the measurement patches, $\lambda_m$, keeping the ratio of measured sites to unmeasured sites fixed, this maximum distance can be tuned larger.  For our simulations, we did not need to perform this measurement patch scaling, because the rate of defect escape was so small, even for $\lambda_m=3$.  Care must be taken, however, because measurement patches cannot be made arbitrarily large without violating the condition mentioned in point \textbf{(2)}. 

 Now, Suppose two defects come into a configuration where they are separated by a distance $C \sqrt{D \te}$ for an integer $C$ and system diffusion constant $D$.  We will estimate the $C$ after which it is just as likely for a defect pair to be corrected by the algorithm as it is to cause an error.  Without loss of generality, define the left defect to be at position $0$.

 Suppose the left defect has recently escaped a measurement patch and been recaptured, thus its estimated age is $0$.  For the error correcting protocol to be able to pair these defects, it must remain on its measurement patch for a time $C\cdot \te$.  But over a timescale $\te$, the defect is equally likely to escape its measurement patch as it is to remain on it, resetting its effective age.  Treating this as a binomial process, we need to estimate the expected amount of time it takes the defect to remain on its measurement patch for $C$ consecutive timescales $\te$.  Call this timescale $\Tau(C,\tau)$ for $C$ consecutive events with timescale $\tau$.  More colloquially---this is equivalent to the expected number of coin flips before a coin has a run of $C$ ``heads'' in a row.  For an event with probability $p$ of occurring, this takes the form

\begin{equation}
\Tau(C,\tau) = \tau \frac{p^{-C} - 1}{1 - p}
\end{equation}
\noindent
Thus, after a time $\Tau(C,\tau)$, a defect could, in principle, be paired with another defect a distance $C \sqrt(D \tau)$ away by the algorithm.

Recall that a diffusion process with constant $D$ will, in $\delta t$ time, become displaced by a distance $\delta x = \sqrt{D \delta t}$.  Because the defects are trapped by the measurement patches, the diffusion rate must be renormalized: $\delta x = \sqrt{D \delta t \cdot (\lambda)/(\gamma_0 \te})$.  This is because it takes a time $\te$ for a defect to actually perform escape from a measurement patch to perform a random step, and this random step has a characteristic length equal to the length of the unit cell, $\lambda$.  Thus, on average, it takes a time $\delta t = C^2 \gamma_0 \te^2 / \lambda$ for a diffusive process to perform an excursion of distance $C \sqrt{D \te}$.

Note that when the timescale over which it takes a defect to remain on a patch $C$ consecutive times equals the timescale it takes a defect to travel a distance $C \sqrt(D \te)$, it is no longer likely for error correction to work.  This only becomes worse as defects become more separated---the pairing defect is more likely to diffuse than it is likely that its pair will remain trapped on a measurement site.  Setting these timescales equal results in the transcendental equation,

\begin{equation}
\te 2^C = \gamma_0 C^2 \te^2 / \lambda
\end{equation}

\noindent Asymptotically for large $\te/\lambda$, $C=\log{(\gamma_0\te/\lambda)}$.  Thus, the lengthscale $\sqrt{D \te} \log{(\gamma_0\te/\lambda)}$ is approximately the maximum correctible lengthscale for our protocol, in the absence of any other corrective measures.  Note that, for our system parameters, this is many thousands of unit cells, and therefore was not detectable by our finite system size analysis.

\subsection{Accurate lifetime estimation}

To ameliorate this issue, we can modify our protocol with additional steps to keep track of defect lifetimes.  That is, within step 1 of the algorithm, perform the following subprotocol:

Let each measurement patch, $m_i$, have two internal clocks, $T^i_1$ and $T^i_2$.

1. If a measurement patch becomes unoccupied without a corrective operation being applied, record the time at which the patch was measured as empty, $T_1=T_{empty}$, and keep the most recent lifetime, $t_{age}$ in memory with a decay constant set by $\tau_{decay}$.  Thus, $T_2 \equiv t_{age}(t) = t_{age_0}\exp(-(t-T_{empty})/\tau_{decay})$.

2. If a defect is subsequently remeasured on this patch at time $t_j$, treat it within the original protocol as if it had been measured at time $t_{age}(t_j)$.  So long as a defect is on the measurement patch, leave $T^j_2$ constant, and update $T_1 = T_{current}$, where $T_{current}$ is the current system clock time.

3. If a defect is subsequently measured on patch $m_i$ with no ``active'' memory of a lifetime---i.e., $T^i_2<.01$---calculate the probabilities given by Eq. \ref{eq:bayes_simple} between $m_i$ and all other unoccupied measurement patches, $m_j$, using $T_{current} - T^j_1$ as the diffusion timescale in Eq. \ref{eq:bayes_simple} for the current system time, $T_{current}$.  Then, probabilistically set $T^i_2$ equal to $T^j_2$, and then reset $T^j_2$ to $0$.

Finally, we impose that $\tau_{decay}$ is several times larger than the diffusive timescale for defects to migrate between measurement patches, but still much smaller than the characteristic timescale over which unpaired defect creation occurs within a unit cell.  This ensures that lifetimes $T^i_2$ decay reasonably quickly if a pair of defects happens to self-annihilate far away from a measurement patch, and will be near $0$ should a new creation event occur, but also ensures that lifetimes are kept in memory long enough to be useful for subsequent re-detection events of escaped defects.

%%%%%%%%%%%%%%%%%%%%%%
\section{Alternative decoding schemes}
\label{sec:appendix_alt_decode_scheme}
%%%%%%%%%%%%%%%%%%%%%%

\begin{figure}
\begin{center}
\scalebox{1}{\includegraphics[width=1.0\columnwidth]{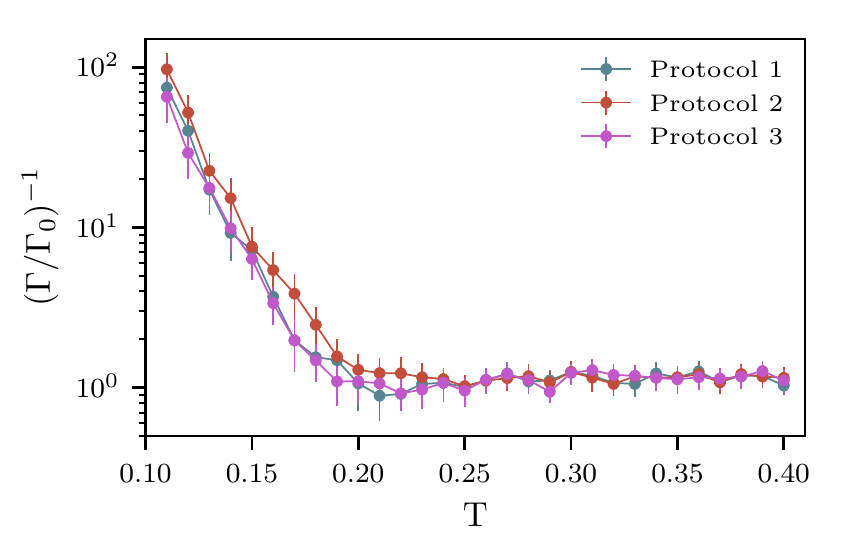}}
\end{center}
\caption{Comparison of three different functional forms used as proxies for $P( d_1:d_2)$.  Protocol 1 is simply the error function expression given by Eq. \ref{eq:bayesformula}.  Protocol 2 is the probability density $\frac{1}{2\pi D |t_2-t_1|} {\rm exp}(-\frac{|x_2-x_1|^2}{2 D |t_2-t_1|})$.  Protocol 3 is the more complicated expression in the last line of Eq. \ref{eq:bayes_simple}.  In practice, each approximation for $P( d_1:d_2)$ is seen to perform approximately equally well.}
\label{fig:decoding_comparison}
\end{figure}

While the approximate Bayesian fusion probability expressed in Eq. \ref{eq:bayes_simple} works well in practice, we find empirically that the precise prefactors of the probability calculation are not terribly important for the decoder functioning correctly.  That is, we find that the final expression in Eq. \ref{eq:bayes_simple} works about as well as $P(d_1^{x_1,t_1} \wedge d_2^{x_2,t_2} | d_1:d_2)$, and that even using the raw probability density,

\begin{equation}
 \frac{1}{2\pi D |t_2-t_1|} {\rm exp}\Bigg(-\frac{|x_2-x_1|^2}{2 D |t_2-t_1|}\Bigg)
\end{equation}

\noindent from Eq. \ref{eq:bayesformula} serves as a decent proxy for the probability, even if this is mathematically dubious in principle.

What is most important for the function of the protocol is that the fusion probability correctly incorporates the expectation that defects are \emph{diffusive}. An amount of sloppiness in this calculation is tolerable, because the defects are efficiently trapped by the protocol, and remain trapped for a long time relative to the diffusive timescales for the system.  But so long as defects that are plausibly ``close'' to one another are the defects that are fused, then we find the protocol to extend system lifetimes effectively.

We plot a comparison of the three aforementioned decoding schemes in Fig. \ref{fig:decoding_comparison}

\bibliography{manuscript_TC_2}

\end{document}